\documentclass{sig-alternate-05-2015}
\usepackage{url,times} 
\usepackage{color}
\usepackage{textcomp}

\usepackage{microtype}
\usepackage{graphicx}
\usepackage{caption}
\usepackage{subfig}
\usepackage{paralist}
\usepackage{cite}
\usepackage[ruled,noline,linesnumbered]{algorithm2e}

\newcommand{\tbd}[1]{\textcolor{red}{XX #1 XX}}

\newcommand{\eat}[1]{}

\newcommand{\mobirnn}{MobiRNN}
\usepackage{microtype}

\usepackage{amsmath}

\def\longconferencenames{}

\newcommand{\conffname}[3]{
\ifx\longconferencenames\undefined
\newcommand{#1}[0]{{#2}}
\else
\newcommand{#1}[0]{{#3}}
\fi
}

\newcommand{\cvnote}[1]{
\ifx\includecvnotes\undefined
\else
{#1}
\fi
}

\newcommand{\bionote}[1]{
\ifx\includebionotes\undefined
\else
{#1}
\fi
}

\conffname{\podc}{PODC}{Proceedings of the ACM symposium on Principles of
distributed computing (PODC)}

\conffname{\asplos}{ASPLOS}{{Proceedings of the ACM International Conference on Architectural Support for Programming Languages and Operating Systems (ASPLOS)}}

\conffname{\spaa}{SPAA}{{Proceedings of the ACM symposium on Parallelism in algorithms and architectures (SPAA)}}

\conffname{\osdi}{OSDI}{{Proceedings of the USENIX Symposium on Operating Systems Design and Implementation (OSDI)}}

\conffname{\disc}{DISC}{{Proceedings of the International Conference on Distributed Computing (DISC)}}

\conffname{\usenixatc}{USENIX}{{Proceedings of the USENIX Annual Technical Conference}}
\conffname{\usenixsec}{USENIX Security}{{Proceedings of the USENIX Security Symposium}}

\conffname{\pldi}{PLDI}{{Proceedings of the ACM SIGPLAN conference on Programming language design and implementation (PLDI)}}

\conffname{\computer}{Computer}{{IEEE Computer}}

\conffname{\sosp}{SOSP}{{Proceedings of the ACM SIGOPS Symposium on Operating Systems Principles (SOSP)}}

\conffname{\isca}{ISCA}{{Proceedings of the ACM IEEE International Symposium on Computer Architecture (ISCA)}}

\conffname{\csaw}{CSAW}{{Proceedings of the ACM Workshop on Computer Security Architecture (CSAW)}}

\conffname{\wddd}{WDDD}{{Proceedings of the Workshop on Duplicating, Deconstructing, and Debunking (WDDD)}}

\conffname{\vldb}{VLDB}{{Proceedings of the International Conference on Very Large Databases (VLDB)}}

\conffname{\toplas}{TOPLAS}{{ACM Transactions on Programming Languages and Systems (TOPLAS)}}

\conffname{\tocs}{TOCS}{{ACM Transactions on Computer Systems (TOCS)}}

\conffname{\ppopp}{{PPoPP}}{{Proceedings of the ACM SIGPLAN Symposium on Principles and Practice of Parallel Programming (PPoPP)}}

\conffname{\jpdc}{J. Parallel Distrib. Comput.}{{Journal of Parallel and Distributed Computing}}

\conffname{\ismm}{ISMM}{{Proceedings of the ACM International Symposium on Memory Management (ISMM)}}

\conffname{\cacm}{CACM}{{Communications of the ACM (CACM)}}

\conffname{\hpca}{HPCA}{{Proceedings of the IEEE International Symposium on High-Performance Computer Architecture (HPCA)}}

\conffname{\transact}{TRANSACT}{{Proceedings of the ACM SIGPLAN Workshop on Transactional Computing (TRANSACT)}}

\conffname{\iiswc}{IISWC}{{Proceedings of the IEEE International Symposium on Workload Characterization (IISWC)}}

\conffname{\tpds}{IEEE Trans, Parallel Distrib. Syst.}{{IEEE Transactions on Parallel and Distributed Systems}}

\conffname{\osr}{OSR}{{ACM Operating Systems Review}}

\conffname{\nsdi}{NSDI}{{Proceedings of the USENIX Symposium on Networked Systems Design and Implementation (NSDI)}}

\conffname{\cc}{CC}{{Proceedings of the International Conference on Compiler Construction (CC)}}

\conffname{\surveys}{ACM Comput. Surv.}{{ACM Computing Surveys}}
\conffname{\icde}{IDCE}{{Proceedings of the IEEE International Conference on Data Engineering (ICDE)}}
\conffname{\fast}{FAST}{{Proceedings of the USENIX Conference on File and Storage Technologies (FAST)}}
\conffname{\eurosys}{{E}uro{S}ys}{{Proceedings of the ACM European Conference on Computer Systems ({E}uro{S}ys)}}
\conffname{\hotos}{HotOS}{{Proceedings of the USENIX Workshop on Hot Topics in Operating Systems (HotOS)}}
\conffname{\oopsla}{OOPSLA}{{Proceedings of the ACM SIGPLAN Conference on Object-Oriented Programming, Systems, Languages, and Applications (OOPSLA)}}
\conffname{\ndss}{NDSS}{{Proceedings of the Network and Distributed System Security Symposium (NDSS)}}
\conffname{\oakland}{Oakland}{{Proceedings of the IEEE Symposium on Security and Privacy (Oakland)}}
\conffname{\ispass}{ISPASS}{Proceedings of the IEEE International Symposium on Performance Analysis of Systems and Software (ISPASS)}

\conffname{\europar}{{E}uro{P}ar}{{Proceedings of the European Conference on Parallel Programming ({E}uro{P}ar)}}

\conffname{\sigcse}{{SIGCSE}}{{Proceedings of the ACM SIGCSE technical symposium on Computer science education (SIGCSE)}}

\conffname{\ccs}{{CCS}}{{Proceedings of the ACM Conference on Computer and Communications Security (CCS)}}

\conffname{\veeconf}{{VEE}}{{Proceedings of the International Conference on Virtual Execution Environments (VEE)}}

\conffname{\lisa}{{LISA}}{{Proceedings of the Large Installation System Administration Conference (LISA)}}
\conffname{\wlpe}{WLPE}{{Workshop on Logic-based methods in Programming Environments (WLPE)}}
\conffname{\acsac}{ACSAC}{{Annual Computer Security Applications Conference (ACSAC)}}
\conffname{\cgo}{{PCGO}}{{Proceedings of the International Symposium on Code Generation and Optimization (CGO)}}
 
\CopyrightYear{2017} 
\setcopyright{acmcopyright} 
\conferenceinfo{EMDL'17,}{June 23, 2017, Niagara Falls, NY, USA}
\isbn{978-1-4503-4962-8/17/06}\acmPrice{\$15.00}
\doi{http://dx.doi.org/10.1145/3089801.3089804}

\title{MobiRNN: Efficient Recurrent Neural Network Execution on Mobile GPU}
\date{}

\def\sharedaffiliation{
\end{tabular}
\begin{tabular}{c}}
\author{Qingqing Cao\and Niranjan Balasubramanian\and Aruna Balasubramanian\\[5pt]
	      \sharedaffiliation
\email{\{qicao, niranjan, arunab\}@cs.stonybrook.edu}\\[5pt]
\affaddr{Stony Brook University}
}

\begin{document}

\maketitle

\begin{abstract}

In this paper, we explore optimizations to run Recurrent Neural Network (RNN) models locally on mobile devices. RNN models are widely used for Natural Language Processing, Machine translation, and other tasks. However, existing mobile applications that use RNN models do so on the cloud. To address privacy and efficiency concerns, we show how RNN models can be run locally on mobile devices. Existing work on porting deep learning models to mobile devices focus on Convolution Neural Networks (CNNs) and cannot be applied directly to RNN models. In response, we present \mobirnn, a mobile-specific optimization framework that implements GPU offloading specifically for mobile GPUs. Evaluations using an RNN model for activity recognition shows that \mobirnn\ does significantly decrease the latency of running RNN models on phones. 

\end{abstract}

\section{Introduction}

Cloud-based deep learning solutions support many mobile applications. The cloud-based applications collect data from the phone, ship it to the cloud, and then apply deep learning models on the cloud. By sending all the data to the cloud, privacy is affected significantly, especially when the data can be used to infer sensitive information such as medical conditions. Cloud connectivity also may not always be available and is often unreliable. 

Due to privacy and reliability concerns, many have explored optimizations for 
running deep learning models locally on the phone, focusing mostly on applications built on
convoluted neural networks (CNNs)~\cite{deepcompression, deepx, deepsense, sparsesep, wu2016quantized, cnncompression}. 
In this work we investigate optimizations for applications built on Recurrent Neural Networks (RNNs), 
a type of deep learning solution common in Natural Language Processing (e.g., Machine Translation) and Human activity recognition. 

Porting RNNs to small form factor devices such as smartphones is a relatively under-studied problem and there are no good solutions. 
First, existing CNN optimizations are not directly useful for RNNs.
The sequential nature of RNNs introduces dependencies that limit the amount of extreme parallelizations.
Second, existing RNN optimizations that perform GPU offloading~\cite{pcgpurnn} are designed for the desktop GPU and do not work well for mobile GPU. Mobile GPUs have fewer GPU cores and a limited integrated memory which limit the benefits of existing offloading techniques. 

In response we develop \mobirnn, a mobile specific optimization for RNNs that focusses on offloading deep learning tasks to the mobile GPU. Our approach to offloading is to use a mobile-specific parallelization framework RenderScript~\cite{renderscript}. At its core, RenderScript offers a way to define computation in terms of some custom data structures. These data structures are parallelized automatically by the RenderScript across the available cores on the GPU. The developer does not make any parallelization decision. 

We evaluate our implementation for the activity recognition task. We use an RNN model that is trained to predict activities based on sensor inputs. The goal of our evaluation is to study the effectiveness of offloading the RNN model to the GPU. Our main result is that \mobirnn\ improves the time to run RNN models by over 4 times; in contrast GPU offloading techniques designed for desktop GPUs {\em degrades} performance by about 4 times. We also find that (1) the speed ups we get with GPU offloading depends on the mobile device type and model complexity, (ii) running a multi-threaded RNN on the CPU gets at least 70\% of the performance benefits that one can get leveraging the GPU, and (iii) an overloaded GPU (a common occurrence in mobile devices since GPU is used for rendering and other tasks) significantly reduces the speed up from GPU offloading.

We open sourced the \mobirnn\ library for researcher's further interest and evaluation. The source code is available at \url{https://github.com/csarron/MobiRNN-EMDL17}.

\section{Background}
\label{sec:background}

Our goal is to study how Recurrent neural networks can be optimized for mobile devices. Much of what is known about mobile device optimizations for deep neural networks exists for Convolutional neural networks (CNNs) and general feed forward deep neural networks (DNNs). In this section, we introduce Long Short-Term Memory (LSTM) models, a widely-used form of Recurrent neural networks and point out the deficiencies in existing optimizations and the gaps in our understanding.

\subsection{Recurrent Neural Networks}

Long Short-Term Memory (LSTM) models, a form of Recurrent neural networks, are highly effective for prediction problems defined over sequential data. As sequential data is fed in, for each time step the models consume one unit of input and make a prediction decision based on both the current input unit and the inputs seen thus far. A key requirement therefore is to remember some information about the past. Often the current decision will depend only on the recent past. For example, in activity recognition, if in the previous steps the user was driving, then it is quite likely that the user is still driving. Sometimes a current decision may depend on an event that happened much earlier in the past, an issue referred to as the long-term dependency problem. Other times the current decision might simply involve erasing what happened in the past as if the sequence is beginning anew. 

LSTMs offer two relevant mechanisms that work together to process input sequence, one symbol at a time. First it provides a memory (cell state), typically a single real-valued vector that maintains state over time. Second it provides various gates (input, forget, and output) that control how this memory is accessed and modified using information in the current input and what is in the memory already. For example, the input gate controls how much the input data can enter the cell while the forget gate decides how should the cell dispose history information and the output gate determines what to let through. Although there are many variants~\cite{lstmodyssey} of LSTM model, we focus on the basic LSTM model~\cite{basiclstm}.

\begin{figure}[th]
	\begin{center}
		\includegraphics[height=0.5\linewidth]{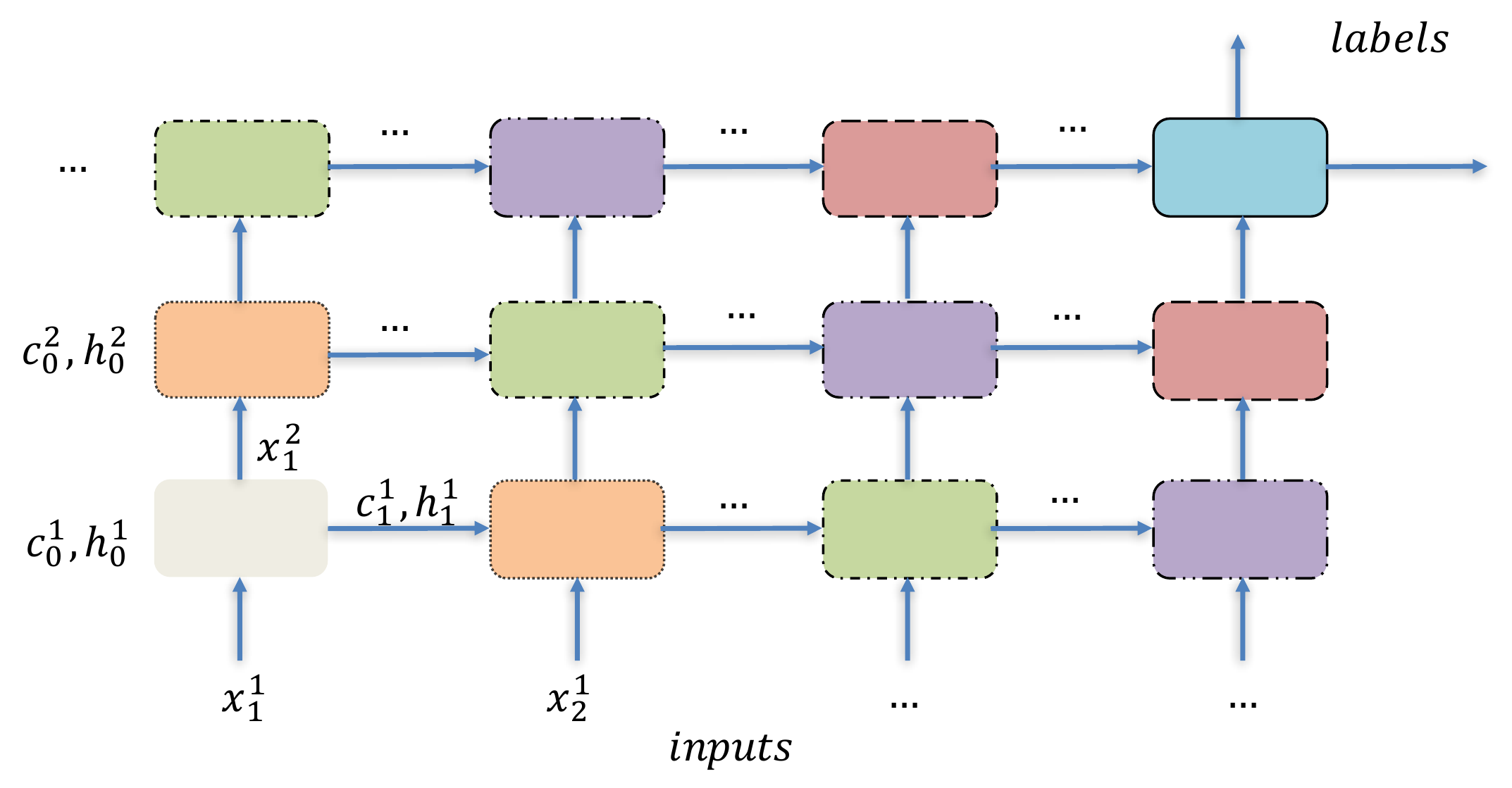}
		\caption{The basic structure of a basic LSTM model. The figure shows the dependencies between the cells.}
		\label{fig:celldepend}
	\end{center}
\end{figure}

Figure~\ref{fig:celldepend} shows the dependencies among the cells as well as the input and output to each cell. Each cell is represented as a rectangle in the figure, and their corresponding inputs and outputs are shown. The input to a cell in layer $i$ at time $t$ is represented as $x_{t}^{i}$, where $x$ is a vector. Each cell also gets input from an earlier cell, represented by vectors $c$ and $h$. $c$ represents the cell state and $h$ represents the hidden state. Both of these inputs are generated by a previous cell. For instance, the input to a cell at (layer $i+1$, time $t+1$) in terms of the cell state, hidden variables, and input $x$ is the output for both a cell at (layer $i$, time $t+1$), and at (layer $i+1$, time $t$). At each time step, the LSTM computes memory updates, and decisions based on the current memory. The computations are realized through matrix multiplications, simple vector additions and a set of non-linear transformations. 

The key difference of the LSTM models with respect to the CNNs or other deep neural networks is that the input is processed sequentially which limits parallelization. As with most deep learning based models the computation involves static pre-learned weight matrices (and biases) and dynamic values (vectors) obtained either from input or from previous computations. However, many computations have a sequential dependency and cannot be fully parallelized unlike models where the entire input is presented at once.

\begin{figure*}[t]
    \centering
    \vspace{-1em}
    \subfloat[][A single operation at an LSTM gate. A vector multiplied with a matrix.]{\includegraphics[scale=0.5]{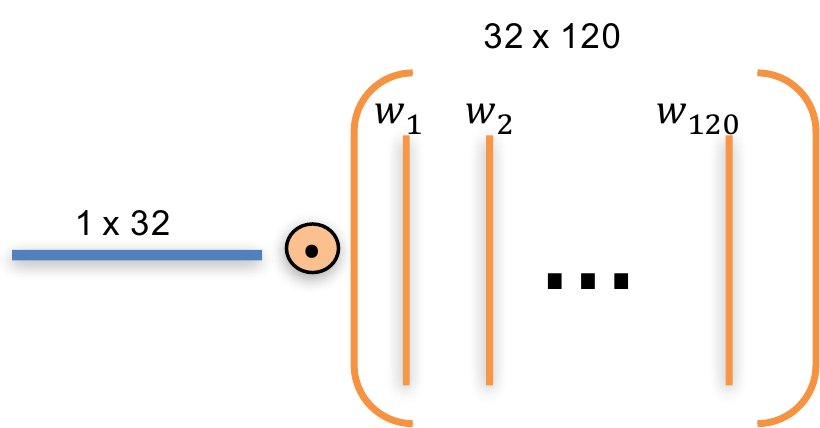}\label{fig:operation}}\hfill
    \subfloat[][CUDA-based offloading for GPU.]{\includegraphics[scale=0.5]{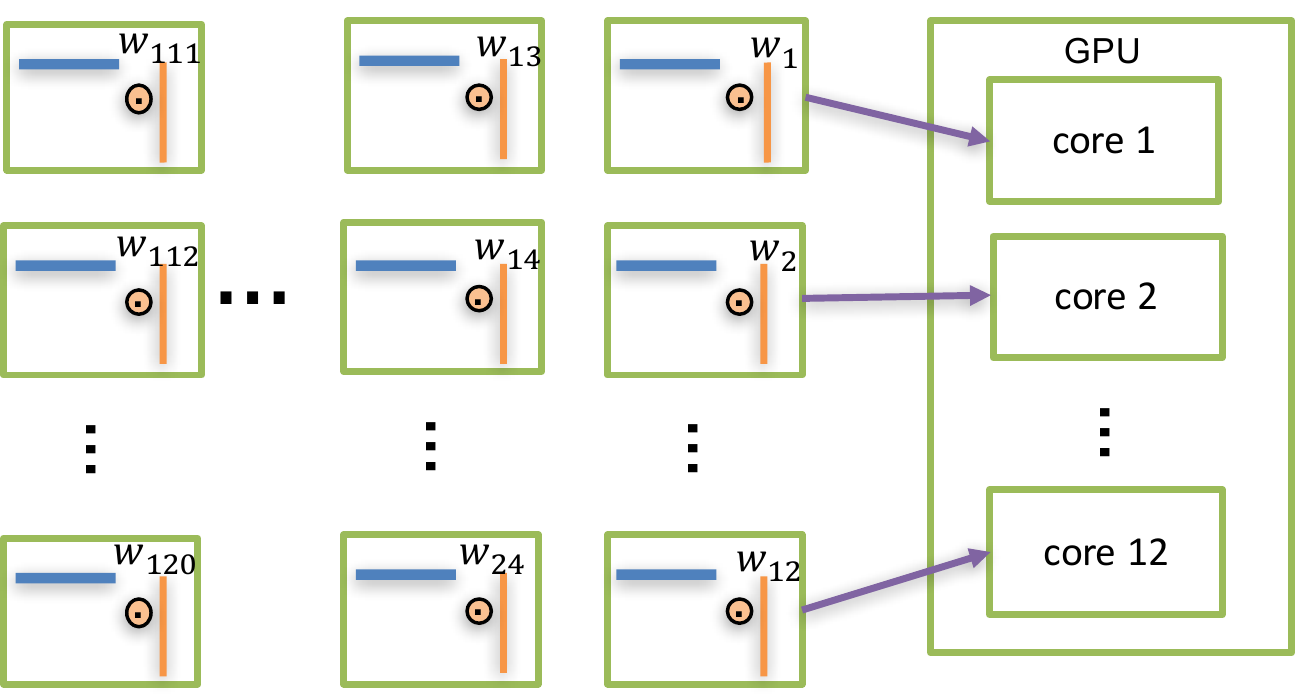}\label{fig:cuda-offloading}}\hfill
    \subfloat[][MobiRNN offloading for GPU]{\includegraphics[scale=0.5]{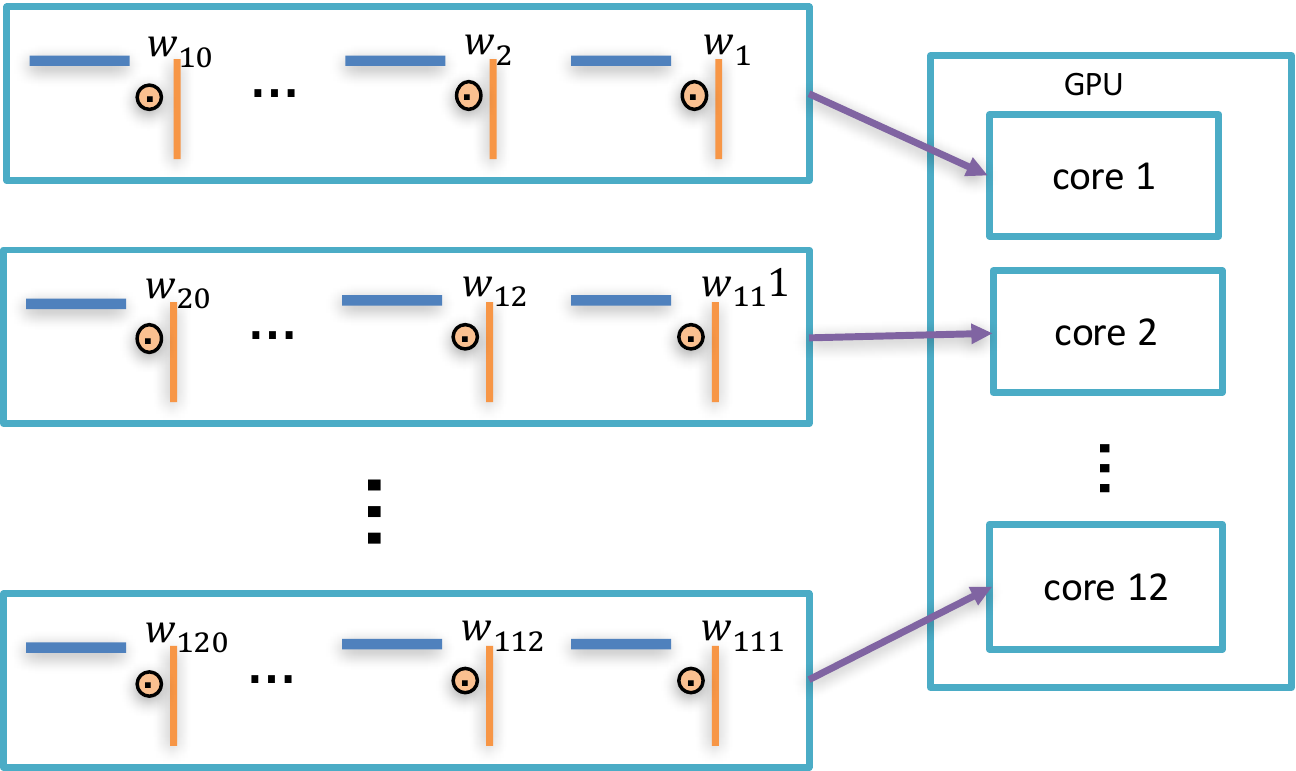}\label{fig:mobirnn-offloading}}
    \caption{\label{fig:adapt-background-size} Adaptation performance compared against background size and number of factors. }
    \vspace{-.5em}
\end{figure*}

\subsection{Existing work on deep learning on mobile devices }
DNN and CNN optimizations for mobile devices focus mostly on compression (reduction model size) and decomposition (parallelizing model computations). Lane et al. pioneered running DNN and CNN on mobile devices ~\cite{deepear, deepx, sparsesep}. DeepEar~\cite{deepear} demonstrated the possibility of applying DNNs to audio sensing area on low-power mobile DSPs. DeepX~\cite{deepx} then enabled the execution of DNN and CNN on mobile devices through runtime layer compression and deep architecture decomposition. Later, SparseSep~\cite{sparsesep} further improved the deep model execution using  a layer compression compiler and adopted sparse inference runtime and convolution separation runtime. 
Because the CNN and DNN models often tend to be large, these compression and decomposition optimizations work especially well. However, for LSTMs that have relatively smaller model sizes, these optimizations are not necessarily effective.

Mobile GPUs provide another optimization avenue. Deep-\break Sense~\cite{deepsense} and CNNDroid~\cite{cnndroid} both showed a mobile GPU can be used to improve the CNN/DNN execution time. For example, in CNNDroid, they reported more than 10-fold speedup for AlexNet model on CIFAR-10 dataset. However, it remains to be seen if the benefits will carryover for LSTMs given the differences in computational dependencies due to their serial architecture. Further, unlike the non-mobile setting, GPUs in the mobile devices are used for other critical applications such as rendering the main UI. It is not clear how much benefit can be squeezed out of the mobile GPU without adversely affecting user experience.

More recent works on RNN(LSTM) optimizations using GPUs~\cite{pcgpurnn} focus on desktop GPUs. But as we describe in the next section, the structure of the desktop GPU is different from the mobile GPU, making it more challenging to port GPU offloading techniques on desktops to mobile devices.

\section{MobiRNN}
\label{sec:impl}

In this work, we dig deep into one optimization technique for neural networks: that of offloading to the GPUs. 

Inherent dependencies between the cells limit the possibility for the kinds of extreme parallelization that are possible in CNNs.
However, there is significant scope for parallelizing operations within each cell. This type of within cell parallelization using a CUDA GPU programming framework has shown performance improvements in desktop CPU settings~\cite{pcgpurnn}.  In this work we show that for the Mobile GPU setting directly applying a CUDA-like model is ineffective and actually deteriorates performance. Here we first describe a direct application of the desktop CUDA model and point out its issues. We then describe our mobile device specific GPU offloading model for RNNs (MobiRNN).

\subsection{CUDA-based GPU Offloading}
The CUDA programming model used in a desktop GPU provides a way to specify how to break down a large unit of computation into smaller work units that then get executed on the GPU~\cite{cuda}. Work units are executed in parallel one in each of the available cores in the GPU. If there are more work units than cores then the units wait until one of the cores becomes available. Figure~\ref{fig:cuda-offloading} illustrates this for computations that are part of a single gate operation -- a 32 dimension input vector multiplied with a (32 x 120) weight matrix. One factorization of this operation is to break this down into a set of 120 vector products, where the input vector multiplied with each of the 120 columns in the matrix. These 120 work units are scheduled twelve at a time leading to 120 function calls to the GPU. 

This type of optimization is ill-suited for mobile GPUs especially when the factorization is not designed carefully. Mobile GPUs have fewer cores and have a shared memory unlike the desktop GPUs. This reduced memory forces a factorization of the work into many small work units. However, this fine-grained factorization incurs a scheduling cost. There is a non-negligible overhead for each work unit, which can quickly add up erasing any gains to be had from parallel execution. 

\begin{figure}[ht]
	\begin{center}
		\includegraphics[width=\columnwidth]{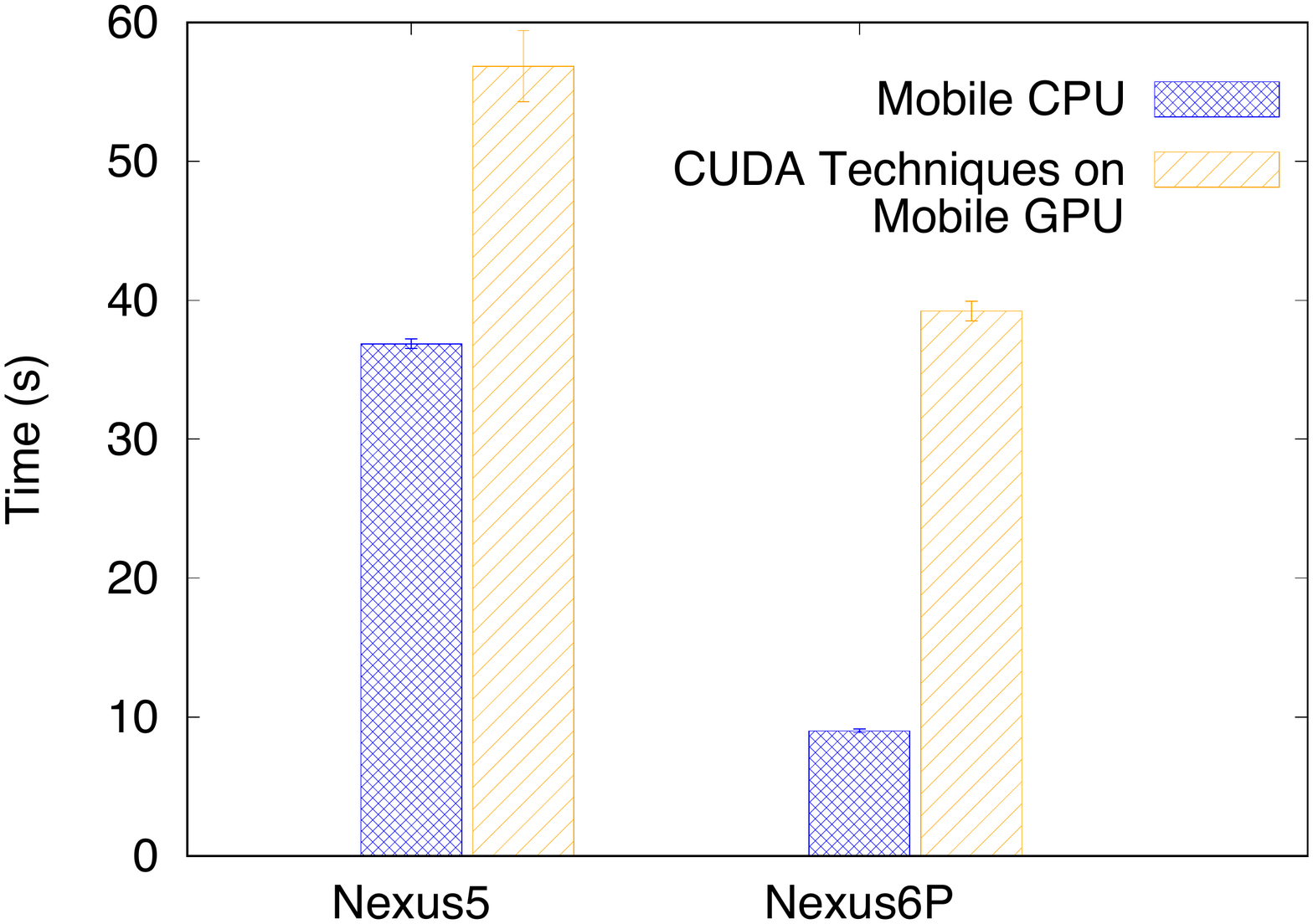}
		\caption{Comparison of offloading the LSTM model to the GPU using the same techniques used on desktop GPUs. The CPU execution time is shown for comparison. Details on the evaluation set up are in \S\ref{sec:eval}.}
		\label{fig:timecuda}
	\end{center}
\end{figure}

Figure~\ref{fig:timecuda} supports this observation. We see that the performance of the LSTM model when run on the GPU using the factorization described for desktop GPUs ~\cite{pcgpurnn}. Details of the LSTM model and the evaluation set up is described later (in \S\ref{sec:eval}). Rather than improving performance, offloading to the GPU made the model run up to 4 times slower compared to running the model on the CPU. 

\eat{
\begin{figure}[t!]
\begin{center}
   \includegraphics[height=0.5\linewidth]{lstm_cell.pdf}
        \caption{The internal structure of an LSTM cell. The computation involves matrix multiplications between the inputs to the cell. Existing techniques that offload this computation to the CPU parallelize each pair of operations between the input vectors.}
        \label{fig:lstmchain}
        \end{center}
\end{figure}
}

\subsection{MobiRNN GPU Offloading}

Our approach to offloading is to use a mobile-specific parallelization framework rather than custom-design work units. 
We use RenderScipt~\cite{renderscript} a high performance language framework that parallelizes work across all processors available on a device, including CPUs and GPUs. At its core, RenderScript offers a way to define computation in terms of some custom data structures designed for data parallel operations, which are then automatically broken down into work units that get executed in the GPU. In this model, the developer makes no factorization decisions. Figure~\ref{fig:mobirnn-offloading} illustrates a type of breakdown that is possible with RenderScript where the vector products are packed into a smaller number of work units -- 12 work units that compute ten vector products each -- thereby reducing the overhead associated with scheduling. 

Further, we also optimize memory allocations for variables using RenderScript primitives that allow for reuse of previously allotted memory, thereby reducing unnecessary and frequent on-demand memory allocation. For example, since the dimension of the cell state($c$) and hidden state($h$) matrix is known as the model is fixed, they can be preallocated. In Figure~\ref{fig:celldepend}, the maximum parallelization is 3, thus only 6(2*3) matrices are preallocated instead of allocating 24(2*3*4), as one cell finishes calculation, the $c$ and $h$ memory are reused.

\subsection{Other Optimizations for MobiRNN}

Divergence statements (if, switch etc) hurt performance in streaming processors like GPU, since they force splits in computation and induce serialization. We carefully architect our code to avoid divergence statements in GPU execution.  
We also use known optimizations like combining inputs and weights, fuse point-wise operations (these two are also suitable for desktop GPUs). 

Other common optimizations like weights quantization, matrix decomposition and  tiling matrix multiplication are not implemented in \mobirnn\ since they are not our focus. The optimization can be further improved by leveraging  hardware features such as using linear algebra library(OpenBlas/Eigen) for matrix operations. Similarly, tilling matrix multiplication used commonly in CUDA-based GEMM can also further improve the performance of MobiRNN.

\section{RNNs on Mobile Devices}
\label{sec:eval}

The goal of our evaluation is to study the effectiveness of offloading deep learning models to the GPU. Different from previous works that focus on Convolutional Neural Networks, our work focuses on Recurrent Neural Network models (in the form of LSTMs). We study the effect of offloading the models to the GPU on mobile devices. Our system \mobirnn\ uses a set of optimization techniques that are well suited for RNNs and for mobile GPU architecture. We implement \mobirnn\ on Android.

Our main findings are that (i) leveraging the GPU for running RNN models does provide speed ups over running the models on the CPU, but the speed up depends on the mobile device and the model complexity, (ii) running a multi-threaded RNN model on a CPU gets at least 70.5\% of the performance benefits that one can get leveraging the GPU, and (iii) an overloaded GPU (a common occurrence in mobile devices since GPU is used for rendering and other tasks) significantly affects performance.

\subsection{Experimental setup and methodology}
\label{sec:setup}

We experiment with an LSTM model that is used for activity recognition~\cite{har}. The stacked LSTM model is trained on the smartphone sensor dataset~\cite{dataset} which has 7352 training and 2947 test sets. For each set, there are 128 readings with 9 dimensions corresponding to the different sensor data. The label data categories each set into one of size activities. The trained model can be used to predict the activity time based on the input sensor data.

We test the model on the mobile device, and train the model on a server. Training is performed using the TensorFlow~\cite{tf} deep learning framework. The model has two parameters that can be tuned: the number of layers and the number of hidden units. In the default case, we configure the model with 2 layers and 32 hidden units. 

 The experiments include running the LSTM model on a randomly selected 100 test cases. The experiments are performed on one Nexus 5 running Android version 6.0.1, and one  Nexus 6p running Android 7.1. 

\subsection{Running LSTM models on the CPU versus GPU}

\begin{figure}[ht]
	\begin{center}
		\includegraphics[width=\columnwidth]{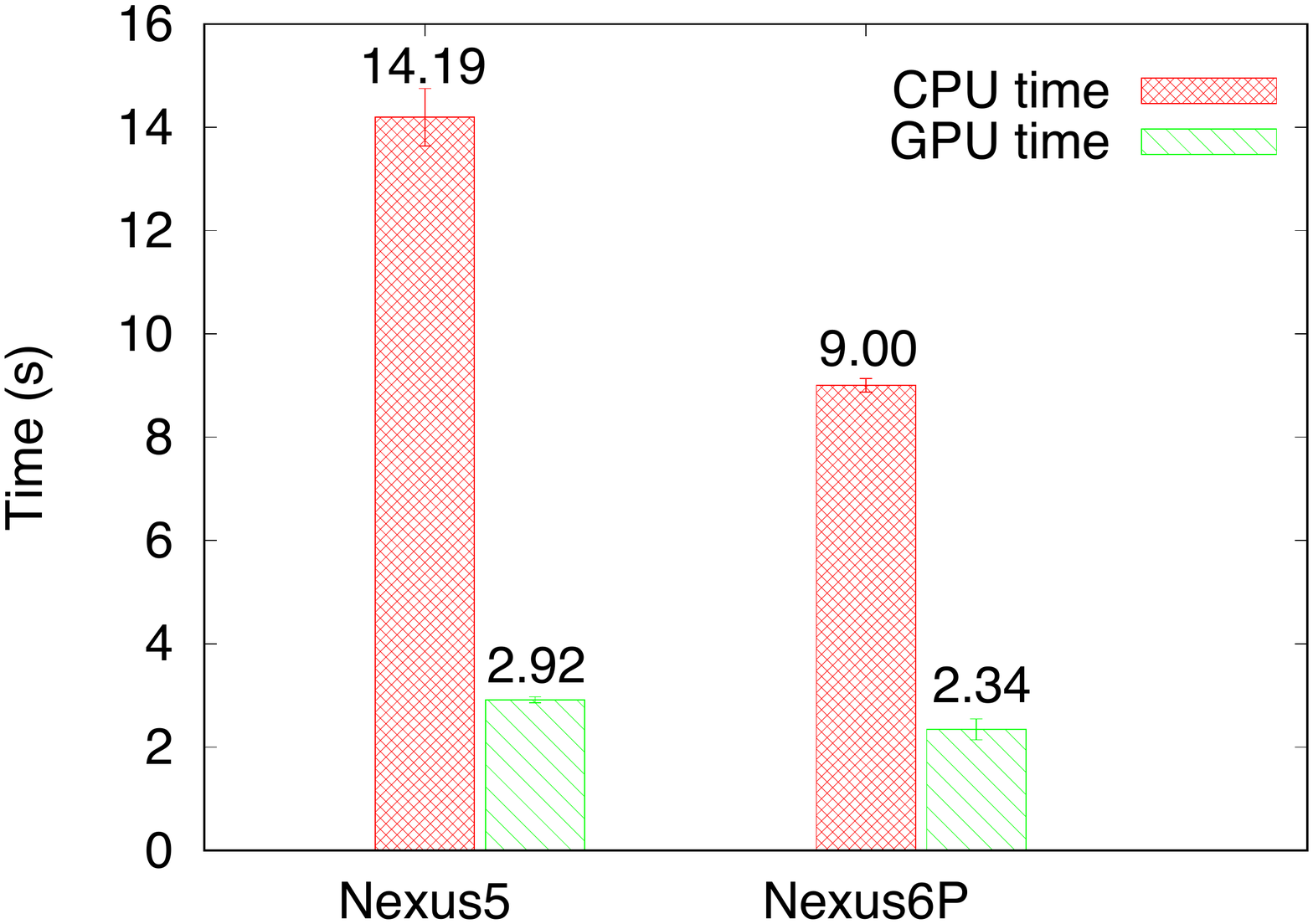}
		\caption{Comparing the relative efficiency of running the LSTM model on the GPU versus the CPU. The results show the aggregate time for running 100 test cases.  }
		\label{fig:timecmp}
	\end{center}
\end{figure}

Figure~\ref{fig:timecmp} shows that \mobirnn\ uses techniques well-suited for mobile GPUs, and as a result, the model runs at least 3.93 times faster on the GPU compared to the CPU. In terms of absolute values, in one case on Nexus 5, the CPU-based classification took 142ms versus 29ms on the GPU.
Recall from Figure~\ref{fig:timecuda} that porting the RNN model to the GPU using CUDA-like techniques performed worse since it was not well-suited for the more constrained mobile GPU. 

However, speed up depends on the phone model. The newer Nexus 6P phones are equipped with a Octa-core CPU (twice the cores of Nexus 5) and higher memory bandwidth (25.6 GB/s, twice of Nexus 5).
Therefore, running the RNN model on the CPU is faster on the Nexus 6P phone compared to Nexus 5.  However, the performance of the RNN model on the GPU are comparable on the two phones. The result is that the speed up we get from running the model on the GPU is 3.93 compared to the 2.83 speed up we get on the newer Nexus 6P phones.

\subsection{Effect of increasing model complexity}

\begin{figure}[ht]
	\begin{center}
		\includegraphics[width=\columnwidth]{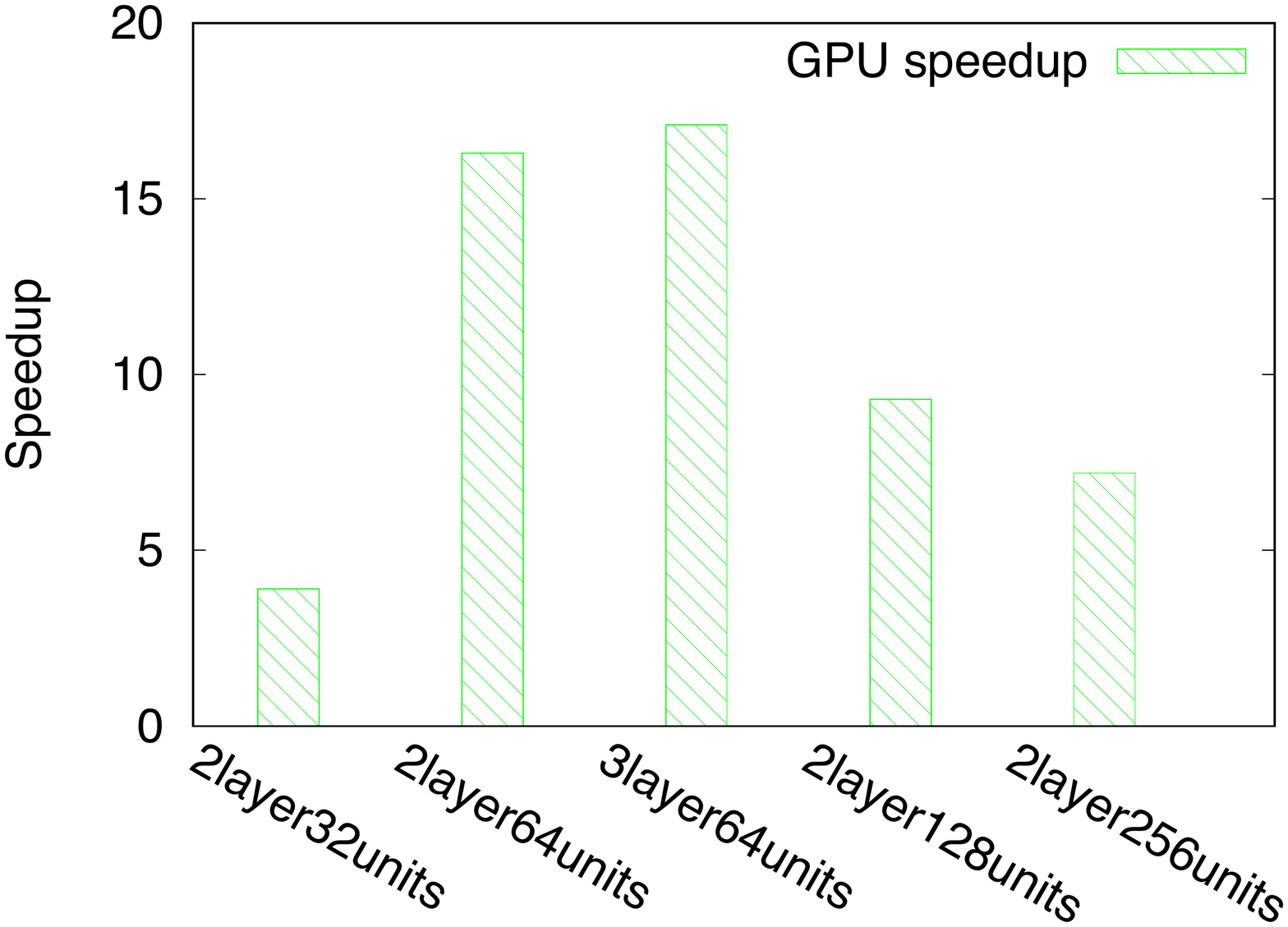}
		\caption{Comparing the relative efficiency of running the LSTM model on the GPU versus the CPU as the model complexity increases. }
		\label{fig:model}
	\end{center}
\end{figure}

Figure ~\ref{fig:model} shows the effect of increasing the model complexity. The model complexity can be increased by either increasing the number of hidden units or the number of layers (\S\ref{sec:background}). On the Nexus 5 phone we increased the number of hidden units from 32 to 256 (the corresponding parameters increase from seventeen thousand to 1 million) and increased the number layers from 1 to 3. 

Figure ~\ref{fig:model} shows the speed up when running the model on GPU using \mobirnn\ compared to running it on the CPU. As the model complexity increases, the speed up using the GPU increases initially. This is to be expected, since for more complex models, parallelization helps even more. 

However, when the model complexity increases due to number of hidden units rather than the number of layers, the speed up due to the use of GPU saturates. Increasing number of hidden layers increases the model size, resulting in the mobile device takes longer to load the parameters. In this case, the memory bandwidth is the bottleneck, decreasing the utility of parallelization.  For example, the LSTM model with 2 layer and 128 hidden units has 263000 parameters, which is four times that of the LSTM model with 2 layer and 64 hidden units. 

\subsection{Multithreading comparison}

Most of the performance benefit from leveraging the GPU comes from parallelization. We evaluate an alternate design point: that of running a multi-threaded LSTM model on the CPU. Porting the LSTM to a multi-threaded CPU is straightforward: when the GPU driver of RenderScript runtime is disabled on the mobile device, the \mobirnn\ GPU implementation simply runs on the CPU using multiple threads. 

\begin{figure}[ht]
	\begin{center}
		\includegraphics[width=\columnwidth]{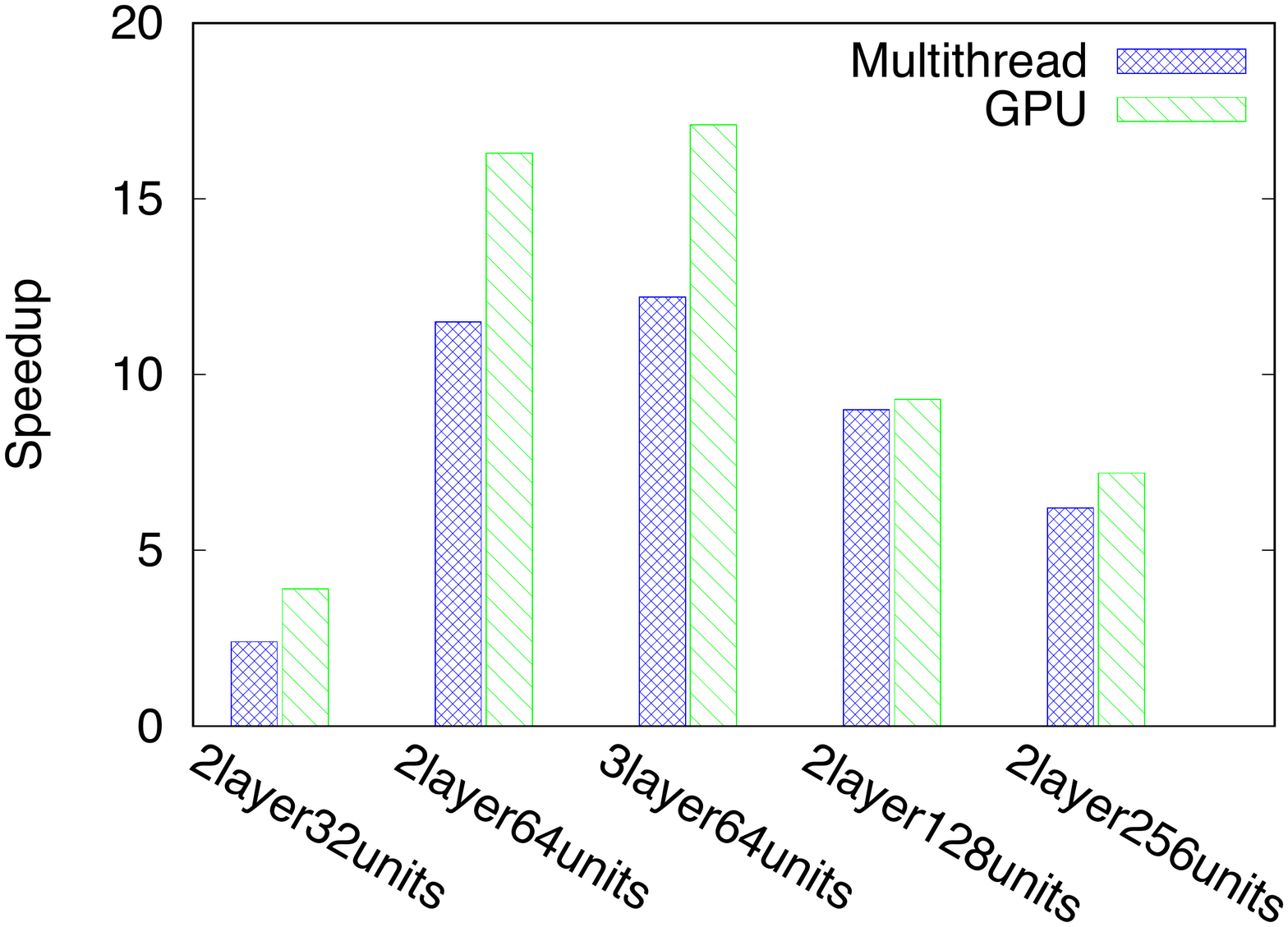}
		\caption{Comparing the performance of running a multi-threaded LSTM model on the CPU compared to porting the model to the GPU.}
		\label{fig:thread}
	\end{center}
\end{figure}

Figure~\ref{fig:thread} shows the speed up when running the LSTM model on the GPU and running the multi-threaded LSTM model on the CPU, for increasing model complexity. These results were obtained on the Nexus 5 phone. Multithreading does speed up performance considerably even when the model is run on the CPU. For one case in the baseline, single thread CPU time is 142ms on average. However, the GPU gives an average of 32\% speed up over the multithreaded version across the models. 

The single threaded CPU version is a stand alone script that does not use RenderScript, whereas both the GPU and the multi-threaded versions use the RenderScript. Part of the speedup over the single threaded version is also because of the efficiency of RenderScript.

We expect that multithreading on the CPU will provide even more benefits on the newer Nexus 6P phone, since the CPU on this newer phone is more powerful. This result shows that the offloading to GPU may not always be the best solution for improving the performance of deep learning models, and other design points need to be explored.

\subsection{Performance of various GPU utilizations}

Unlike dedicated GPU engines, the Android GPU is designed to perform a number of tasks. For example, on Android 3.0~\cite{ha} and above, screen rendering is performed on the GPU since it supports hardware acceleration. If the GPU is overloaded, then offloading deep learning tasks to the GPU may likely affect performance considerably.

To this end, we evaluated \mobirnn\ under different GPU load on the Nexus 6P phone. We perform these experiments under three GPU loads: low utilization ($<$30\%), medium (30$\sim$50\%) and high($>$70\%) load. To obtain the GPU utilization, we use ADB scripts, but for GPUs such as Adreno, the GPU utilization can be obtained using APIs~\cite{qdn}. 

\begin{figure}[ht]
	\begin{center}
		\includegraphics[width=\columnwidth]{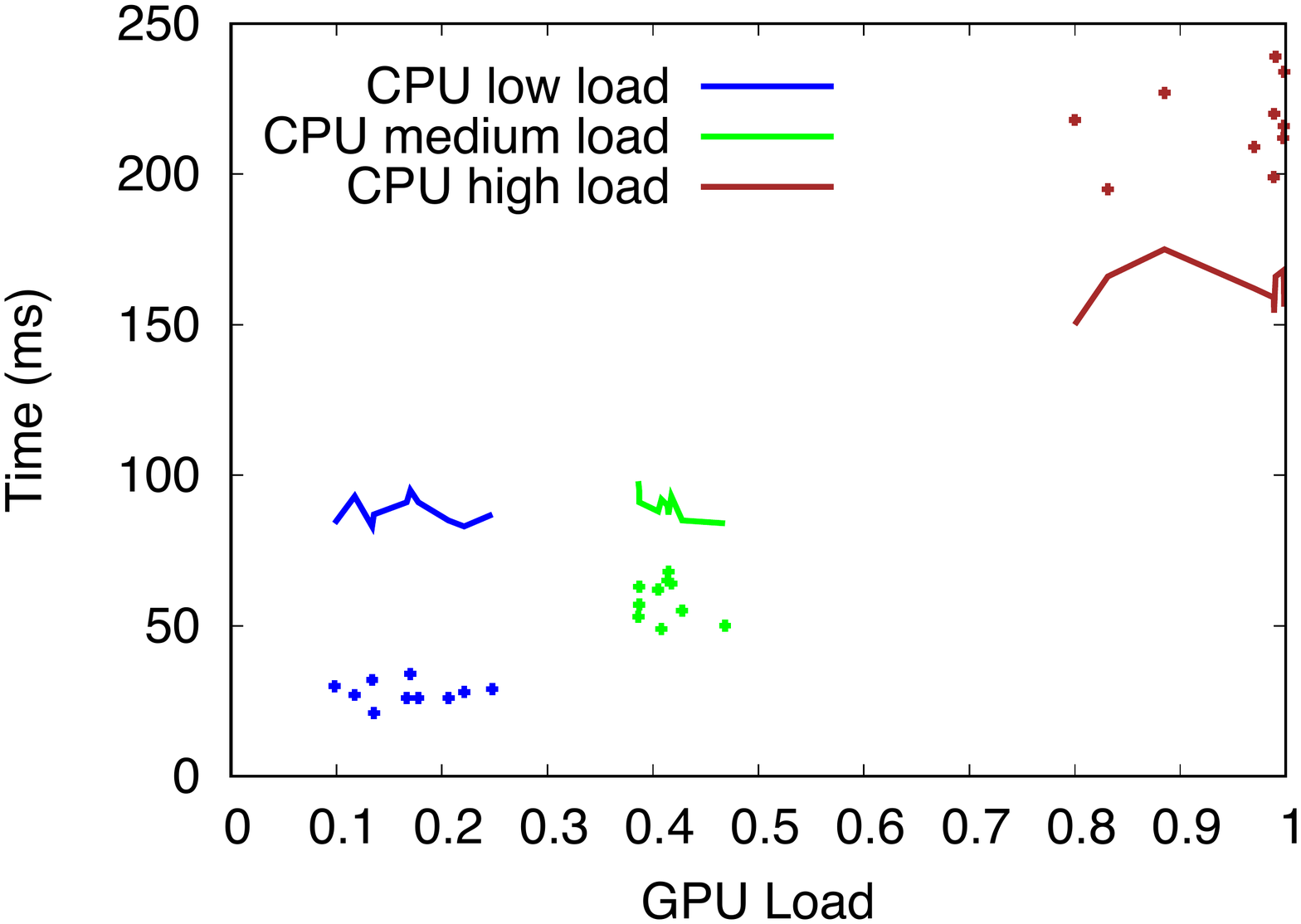}
		\caption{Time to run LSTM on as load on the processor increases. Dots show time when running LSTMs on GPU. Lines show performance when running LSTMs on CPU. }
		\label{fig:util}
	\end{center}
\end{figure}

Figure ~\ref{fig:util} shows that the latency of the LSTM model correlates with the load on the GPU; as the GPU is more loaded, it takes longer for the LSTM model to run on the CPU.  GPU load is correlated with the time it takes to run the LSTM model. As the load on the GPU increases, the time it takes to run the model also increases. For comparison, we show the time taken to run the same LSTM models on the CPU.  For fairness in comparison, we perform these experiments under similar low/medium/high CPU loads. We find that under low and medium loads, running the LSTM model on the GPU reduces latencies; however, under both high GPU and high CPU load it is better to run the RNN on the CPU rather than the GPU. This result suggests that \mobirnn\ should take into account GPU utilization before offloading tasks to the GPU. 

\section{Conclusion}

In this work, we show how GPU offloading can be used to optimize Recurrent Neural Network models on mobile devices. Porting GPU offloading techniques designed for desktop GPUs {\em as-is} to mobile devices performs poorly because the mobile GPUs are much more constrained. Instead, we design a set of mobile-specific techniques for GPU offloading on mobile devices that we call \mobirnn. Our evaluations show that \mobirnn\ does significantly reduce the time to run RNN models on mobile devices. However, the speed up due to GPU offloading depends on the mobile device type, model complexity, and the load on the GPU.

\bibliographystyle{abbrv}
\bibliography{ref}

\end{document}